\documentclass{emulateapj}
\usepackage{apjfonts}
\usepackage{xspace}
\usepackage{amsmath}
\usepackage{mathtools}
\usepackage{hyperref} 
\hypersetup{%
    colorlinks=false, 
    urlcolor=magenta, 
    linkcolor=blue, 
    citecolor=blue,
    breaklinks=true, 
    bookmarksnumbered=true,
    bookmarksopen=false, 
    bookmarksopenlevel=0,
    hypertexnames=true, 
    unicode=true,          
    pdftoolbar=true,        
    pdfmenubar=true,        
    pdffitwindow=false,     
    pdfstartview={FitH}    
}

\usepackage{etoolbox}
\makeatletter

\patchcmd{\NAT@citex}
  {\@citea\NAT@hyper@{\NAT@nmfmt{\NAT@nm}\NAT@date}}
  {\@citea\NAT@nmfmt{\NAT@nm}\NAT@hyper@{\NAT@date}}
  {}
  {}

\patchcmd{\NAT@citex}
  {\@citea\NAT@hyper@{%
     \NAT@nmfmt{\NAT@nm}%
     \hyper@natlinkbreak{\NAT@aysep\NAT@spacechar}{\@citeb\@extra@b@citeb}%
     \NAT@date}}
  {\@citea\NAT@nmfmt{\NAT@nm}%
   \NAT@aysep\NAT@spacechar%
   \NAT@hyper@{\NAT@date}}
  {}
  {}

\patchcmd{\NAT@citex}
  {\@citea\NAT@hyper@{%
     \NAT@nmfmt{\NAT@nm}%
     \hyper@natlinkbreak{\NAT@spacechar\NAT@@open\if*#1*\else#1\NAT@spacechar\fi}%
       {\@citeb\@extra@b@citeb}%
     \NAT@date}}
  {\@citea\NAT@nmfmt{\NAT@nm}%
   \NAT@spacechar\NAT@@open\if*#1*\else#1\NAT@spacechar\fi%
   \NAT@hyper@{\NAT@date}}
  {}
  {}

\makeatother

\begin{document}
\shorttitle{Predicting Merger-Induced Gas Motions in Galaxy Clusters}
\slugcomment{The Astrophysical Journal, accepted}
\shortauthors{Nagai et al.}

\title{Predicting Merger-Induced Gas Motions in $\Lambda$CDM Galaxy Clusters}
\author{Daisuke Nagai\altaffilmark{1,2,3}}
\author{Erwin T. Lau\altaffilmark{1,2}}
\author{Camille Avestruz\altaffilmark{1,2}}
\author{Kaylea Nelson\altaffilmark{3}}
\author{Douglas H. Rudd\altaffilmark{1,2,4,5}}

\affil{ 
$^1$Department of Physics, Yale University, New Haven, CT 06520, U.S.A.; \href{mailto:daisuke.nagai@yale.edu}{daisuke.nagai@yale.edu} \\
$^2$Yale Center for Astronomy \& Astrophysics, Yale University, New Haven, CT 06520, U.S.A. \\
$^3$Department of Astronomy, Yale University, New Haven, CT 06520, U.S.A. \\
$^4$Kavli Institute for Cosmological Physics, University of Chicago, Chicago, IL 60637, USA \\
$^5$Research Computing Center, University of Chicago, Chicago, IL 60637, USA
}


\begin{abstract}
In the hierarchical structure formation model, clusters of galaxies form through a sequence of mergers and continuous mass accretion, which generate significant random gas motions especially in their outskirts where material is actively accreting. Non-thermal pressure provided by the internal gas motions affects the thermodynamic structure of the X-ray emitting intracluster plasma and introduces biases in the physical interpretation of X-ray and Sunyaev-Zeldovich effect observations. However, we know very little about the nature of gas motions in galaxy clusters. The ASTRO-H X-ray mission, scheduled to launch in 2015, will have a calorimeter capable of measuring gas motions in galaxy clusters at the level of $\lesssim$100~km/s. In this work, we predict the level of merger-induced gas motions expected in the $\Lambda$CDM model using hydrodynamical simulations of galaxy cluster formation. We show that the gas velocity dispersion is larger in more massive clusters, but exhibits a large scatter. We show that systems with large gas motions are morphologically disturbed, while early forming, relaxed groups show a smaller level of gas motions. By analyzing mock ASTRO-H observations of simulated clusters, we show that such observations can accurately measure the gas velocity dispersion out to the outskirts of nearby relaxed galaxy clusters. ASTRO-H analysis of merging clusters, on the other hand, requires multi-component spectral fitting and enables unique studies of substructures in galaxy clusters by measuring both the peculiar velocities and the velocity dispersion of gas within individual sub-clusters.
\end{abstract}

\section{Introduction}

Clusters of galaxies are the largest gravitationally bound objects and are formed both through mergers and continuous mass accretion.  These merging and accretion events generate a significant level of gas motions inside the cluster potential well, which eventually heat the gas through shocks or turbulent dissipation.  Cosmological hydrodynamical simulations predict that such gas motions are ubiquitous in galaxy clusters forming in the $\Lambda$CDM model and that a non-negligible ($5-30$\%) fraction of the total energy content of the intracluster medium (ICM) is in the form of kinetic energy \citep{vazza_etal09,lau_etal09}.

The total energy content of the ICM is a useful quantity in cluster-based cosmology as it directly relates to cluster mass. Observations of the hot ICM with X-ray and the Sunyaev-Zeldovich (SZ) effect are, however, sensitive to the thermal energy content only.  Failing to consider the gas kinetic energy fraction can result in a number of systematic biases. For example, the increasing kinetic energy fraction at larger radii leads to suppressed temperatures and flatter entropy profiles in cluster outskirts \citep[][for a review]{reiprich_etal13}. Gas motions also contribute to the support against gravity which leads to biases in cluster mass estimates based on the assumption of hydrostatic equilibrium \citep[e.g.,][]{rasia_etal06,nagai_etal07b,piffaretti_etal08}, and is currently one of the dominant sources of systematic uncertainty in the calibration of cluster observable-mass relations and cosmological parameters derived from clusters \citep{allen_etal08,vikhlinin_etal09,planck_XX13}.  A large kinetic energy fraction in cluster outskirts also reduces the thermal SZ signal of individual clusters and the SZ fluctuation power spectrum \citep{shaw_etal10,battaglia_etal10,trac_etal11}, introducing additional uncertainties in cosmological inference from ongoing SZ surveys \citep{reichardt_etal12,sievers_etal13}.  In addition, gas motions are thought to be responsible for dispersing metals throughout the ICM via turbulent mixing \citep[e.g.,][]{rebusco_etal05} and accelerating particles which gives rise to radio halos in clusters \citep[e.g.,][]{brunetti_etal07}. 

Despite the important role of gas motions in cluster astrophysics and cosmology, we know very little about them observationally.  Indirect evidence from \emph{Chandra}/\emph{XMM}-Newton observations of surface brightness fluctuations in the central regions of the nearby Coma cluster suggest that $\approx 10$\% of ICM energy is in the form of gas motions \citep{schuecker_etal04,churazov_etal12}. There are also constraints from the lack of detection in the shifting and broadening of heavy ion lines \citep[e.g.,][]{ota_etal07}. Although a merging sub-cluster has recently been detected through the line shift \citep{tamura_etal11}, the line broadening has not yet been detected, indicating that turbulent gas velocities are smaller than several hundred km/s on the scale of a few hundred kpc in the central regions of galaxy clusters \citep{sanders_etal13}.  Moreover, no observation to date has provided meaningful constraints on gas motions at large radii where numerical simulations predict they should be significant.

The ASTRO-H mission \citep{takahashi_etal10}, a joint Japanese-US X-ray telescope scheduled to be launched in 2015, has a unique capability to measure gas motions in clusters. The Soft X-ray Spectrometer (SXS) onboard ASTRO-H will have an energy resolution of approximately $7$~eV, which enables measurements of the peculiar velocity and internal gas motions in clusters directly for the first time through the shifting and broadening of Fe lines. It will have spatial resolution of 1.3 arcmin, and the field-of-view of $3.05\times 3.05$ arcmin$^2$. Despite the limited angular resolution and field-of-view, it should be possible to study the velocity structure of the ICM over a wide range of spatial scales ($\approx 1-1000$~kpc) by measuring the 3D gas velocity power spectrum \citep{zhuravleva_etal12} and resonant scattering of the Fe XXV line at $6.7$~keV \citep{zhuravleva_etal11}\footnote{In the presence of gas turbulence, the effect of resonant scattering is suppressed, decreasing the optical depth of resonant scattering line e.g. the Fe XXV line at $6.7$~keV.}. 

Theoretically, several works have investigated the properties of internal gas motions in galaxy clusters using hydrodynamical simulations. It has been suggested that measurements of gas velocity structure may provide insights into the baryonic physics in cluster central regions.  For example, radiative cooling and star formation induce rotational gas motions in the cluster center \citep{rasia_etal04,fang_etal09,lau_etal11,biffi_etal11}. Measurements of the gas velocity anisotropy might help constrain the magnitude of the baryon dissipation \citep{lau_etal12, bianconi_etal13}. \citet{vazza_etal06,vazza_etal11} studied how the kinetic energy fraction in clusters depends on cluster mass and their dynamical state using a sample of $21$ and $20$ simulated clusters, respectively. By analyzing mock spectra of 43 simulated clusters, \citet{biffi_etal13} suggested that the gas velocity measurements with future X-ray missions can identify disturbed clusters and might help reduce the scatter in cluster scaling relations by excluding systems with high gas velocity dispersion. 

The goal of this work is to characterize the gas motions in the virialized regions of galaxy clusters induced by mergers and accretion events and study their detectability with the upcoming ASTRO-H mission.  In this work, we will use a large ($\gtrsim 400$) volume-limited sample of simulated galaxy clusters and groups in order to characterize the range of internal gas motions for a wide range of cluster masses and dynamical states.  Since the {\em total} kinetic energy is the quantity directly relevant for the cluster's energy budget, we are primarily interested in the sum of {\em both} bulk and turbulent motions induced by mergers and accretion. Therefore, our strategy is to focus on non-radiative simulations, in which gas motions are generated by gravitational physics and hydrodynamics of gas alone. We will then assess the roles of non-gravitational physics, such as gas cooling, star formation, and heating by stars and black holes, using a smaller set of cluster simulations that follow these processes self-consistently.  We investigate the effects of baryonic physics (e.g., radiative cooling, star formation, and energy injection from stars and black holes) on gas velocity structure, focusing on the sphere of influence of AGN feedback and the relative importance of the merger vs. AGN induced gas flows.  By analyzing mock ASTRO-H spectra of simulated clusters, we will show that ASTRO-H observations are capable of measuring the internal gas motions in clusters out to $r\approx r_{500c}$ as well as enabling a unique study of cluster gas substructures. 

This paper is organized as follows. Section~\ref{sec:sim} describes numerical simulations and mock ASTRO-H simulations used in this work.  Results are presented in Section~\ref{sec:results}. We provide a summary and discuss implications of our results in Section~\ref{sec:summary}.

\section{Simulations}
\label{sec:sim}

\subsection{Hydrodynamical Simulations}

The simulations presented in this work are performed using the Adaptive Refinement Tree (ART) $N$-body$+$gas-dynamics code \citep{kra99,kra02,rudd_etal08}, which is an Eulerian code that uses adaptive refinement in space and time, and non-adaptive refinement in mass \citep{klypin_etal01} to achieve the dynamic ranges necessary to resolve the cores of halos formed in self-consistent cosmological simulations. The code is parallelized using Massage Passing Interface (MPI) libraries and OpenMP directives.

To study the variation of gas velocity structure with cluster formation history, we analyze a volume-limited sample of $458$ galaxy groups and clusters with $M_{500c}\ge 5\times 10^{13}h^{-1}M_{\odot}$ selected from a $250h^{-1}$~Mpc volume with the same initial conditions as the Bolshoi Simulation \citep{klypin_etal11}, but performed with non-radiative hydrodynamics (Bolshoi NR, hereafter).  This simulation assumes a flat {$\Lambda$}CDM model with {\it WMAP} five-year (WMAP5) cosmological parameters: $\Omega_{\rm m}=1-\Omega_{\Lambda}=0.27$, $\Omega_{\rm b}=0.0469$, $h=0.7$ and $\sigma_8=0.82$, where the Hubble constant is defined as $100h{\ \rm km\ s^{-1}\ Mpc^{-1}}$, and $\sigma_8$ is the mass variance within spheres of radius $8h^{-1}$~Mpc.  The simulation was run using a uniform $512^3$ grid with 8 levels of mesh refinement, implying a maximum spatial resolution of $3.8h^{-1}$~kpc. The simulation is performed with $1024^3$ particles spread uniformly throughout the box, corresponding to dark matter mass resolution of $1.08\times 10^9h^{-1}M_{\odot}$. 

To investigate the effects of baryonic physics on the gas velocity structure, we analyzed a sample of high-resolution hydrodynamical simulations of galaxy clusters formation from \citet[][hereafter N07]{nagai_etal07a,nagai_etal07b}\footnote{Note that the cosmological parameters adopted here differ from the Bolshoi simulations.}, which were simulated with varying physical processes. The first set is performed with non-radiative (NR) gas physics, as with Bolshoi NR.  The second set includes radiative cooling, star formation, metal enrichment, and stellar feedback (CSF). Comparisons of the NR and CSF runs to \emph{Chandra} X-ray cluster observations have been presented in \citet{nagai_etal07a}, illustrating that these two runs bracket observed clusters in a number of properties. The third set includes CSF and energy feedback from supermassive black holes (CSF+AGN; see Avestruz et al., in prep. for more details). Briefly, black hole (BH) particles are seeded with an initial mass of $10^{5}h^{-1}M_{\odot}$ at the centers of dark matter halos with $M_{500c}\gtrsim 2\times 10^{11}h^{-1}M_{\odot}$.  Throughout cosmic history, these BH particles accrete gas with a rate given by a modified Bondi accretion model with a boost parameter \citep{booth_etal09} and return a fraction of the accreted rest mass energy into the environment in the form of thermal energy. 

In particular, we focus on two X-ray luminous clusters, CL104 and CL101, at $z=0$, with the core-excised X-ray temperature of $T_{\rm X}=7.7$ and $8.7$~keV, respectively. CL101 is a massive, dynamically active cluster, which has recently experienced violent mergers (at $z\sim 0.1$ and $z\sim 0.25$) and contains two major substructures near the core at $z=0$. These two substructures have been identified by visual inspection and masked out before its analysis. CL104 is a similarly massive cluster, but with a more quiescent mass accretion history. No significant mergers occur in the last 8 Gyrs of the cluster's mass accretion history, making it one of the most relaxed systems in the N07 sample. Each cluster is simulated using a $128^3$ uniform grid with 8 levels of refinement. Clusters are selected from 120$h^{-1}$~Mpc computational boxes, achieving peak spatial resolution of $\approx 3.6h^{-1}$~kpc, sufficient to resolve dense gas clumps in the ICM. The dark matter particle mass in the region surrounding the cluster is $9\times 10^8 h^{-1}M_{\odot}$, while the rest of the simulation volume is followed with lower mass and spatial resolution. We refer readers to N07 for further details. 

\subsection{Mock ASTRO-H Simulations}

We generate a mock ASTRO-H surface brightness map and spectrum for each orthogonal projection for each cluster in the sample. The ASTRO-H pipeline consists of two main steps: (1) generation of the flux map from the simulation output; and (2) conversion of the flux map to a photon map by convolving with the instrumental response of ASTRO-H. For the first step, we compute the emissivity $\epsilon_E$ as a function of energy $E$ for each hydrodynamic cell using the APEC plasma code \citep{smith_etal01} with AtomDB version 2.0.2 \citep{foster_etal12}.  The emissivity $\epsilon_E = \epsilon_E(\rho,T,Z,z_{\rm obs}, v_{\rm los})$ is a function of the gas density $\rho$, temperature $T$, the gas metallicity $Z=0.3Z_\odot$, observed redshift of the clusters $z_{\rm obs}=0.068$\footnote{This is the redshift of A1795, a nearby relaxed cluster, which is one of the likely targets for mapping out the gas velocity structure out to large radii with ASTRO-H.}, and line-of-sight velocity $v_{\rm los}$ of the cell. The energy range is $E \in [5.0,10.0]$~keV with energy resolution of $\Delta E=1$~eV, and we include the effect of thermal broadening on the emission line. 

Each flux map is computed by summing up the emissivity of all cells contained within a sphere of radius of $5h^{-1}$~Mpc centered on the densest dark matter particle.  We then convolve our flux map with the ASTRO-H Auxiliary Response File (ARF) ({\tt sxt-s\_120210\_ts02um\_intall.arf}) and Redistribution Matrix File (RMF) ({\tt ah\_sxs\_7ev\_basefilt\_20100712.rmf}) response files from the ASTRO-H website.\footnote{\url{http://astro-h.isas.jaxa.jp/researchers/sim/response.html}} The energy resolution for these response file is $7$~eV. We then draw photons for each location from the convolved flux map assuming a Poisson distribution.  The number of photons depends on our chosen exposure time $t_{\rm exp}$, which is a free parameter. 

For each mock ASTRO-H map, we extract a spectrum from the regions of interest. We then measure both peculiar velocity and velocity dispersion of gas by performing spectral fitting of the the mock ASTRO-H spectrum using XSPEC version 12.8, using the BAPEC model with temperature, metal abundance, redshift, velocity dispersion, and the spectrum normalization as free parameters, in the energy range of $E \in [6.05,6.95]$~keV. Thermal broadening is included when fitting for the emission lines. The fitting is performed using Cash statistics \citep{cash79}. 

\begin{figure}[t]
\begin{center}
\epsscale{1.2}\plotone{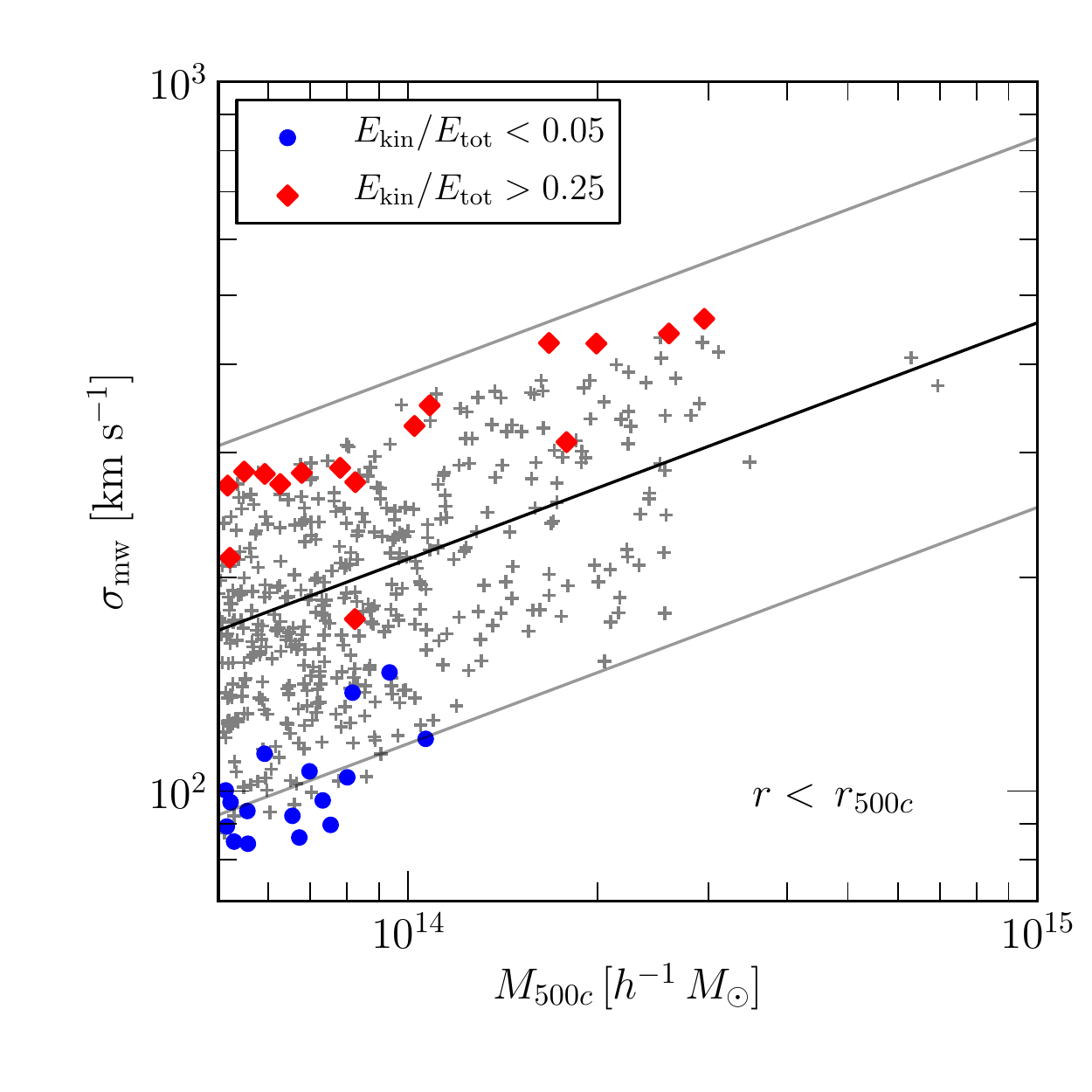}
\caption{One dimensional velocity dispersion of gas within a sphere of $r_{500c}$ as a function of $M_{500c}$.  The points indicate all clusters with $M_{500c}>5\times 10^{13}h^{-1}M_{\odot}$.  The red diamonds and blue circles indicate clusters with the kinetic energy fraction of $E_{\rm kin}/E_{\rm tot}>0.25$ and $<0.05$, respectively.  The middle line indicates the best-fit relation with fixed slope of $1/3$, and the two surrounding lines indicate 2$\sigma$ around the fit. 
}
\label{fig1}
\end{center}
\end{figure}

\begin{figure}[t]
\begin{center}
\epsscale{1.2}\plotone{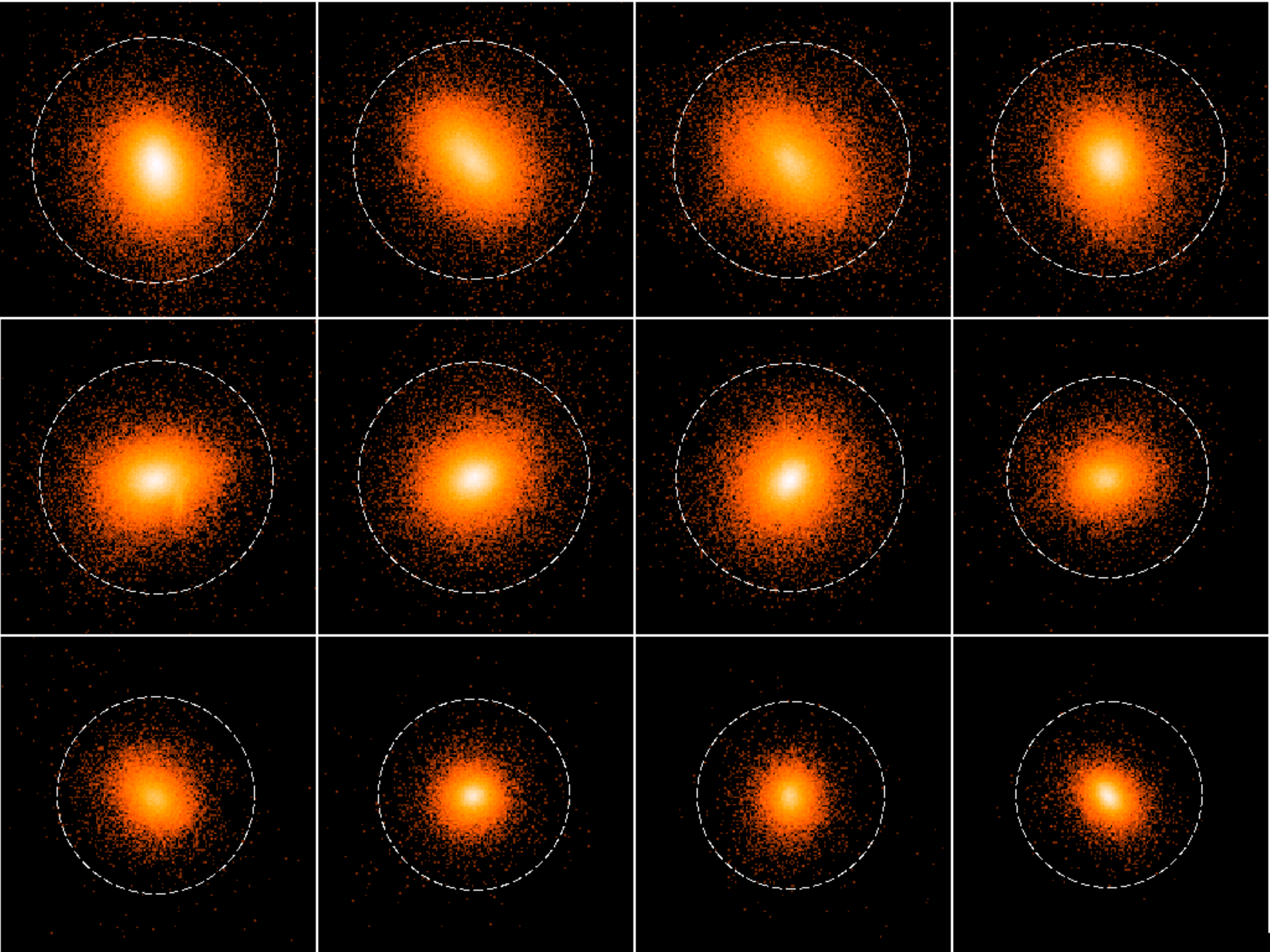}
\epsscale{1.2}\plotone{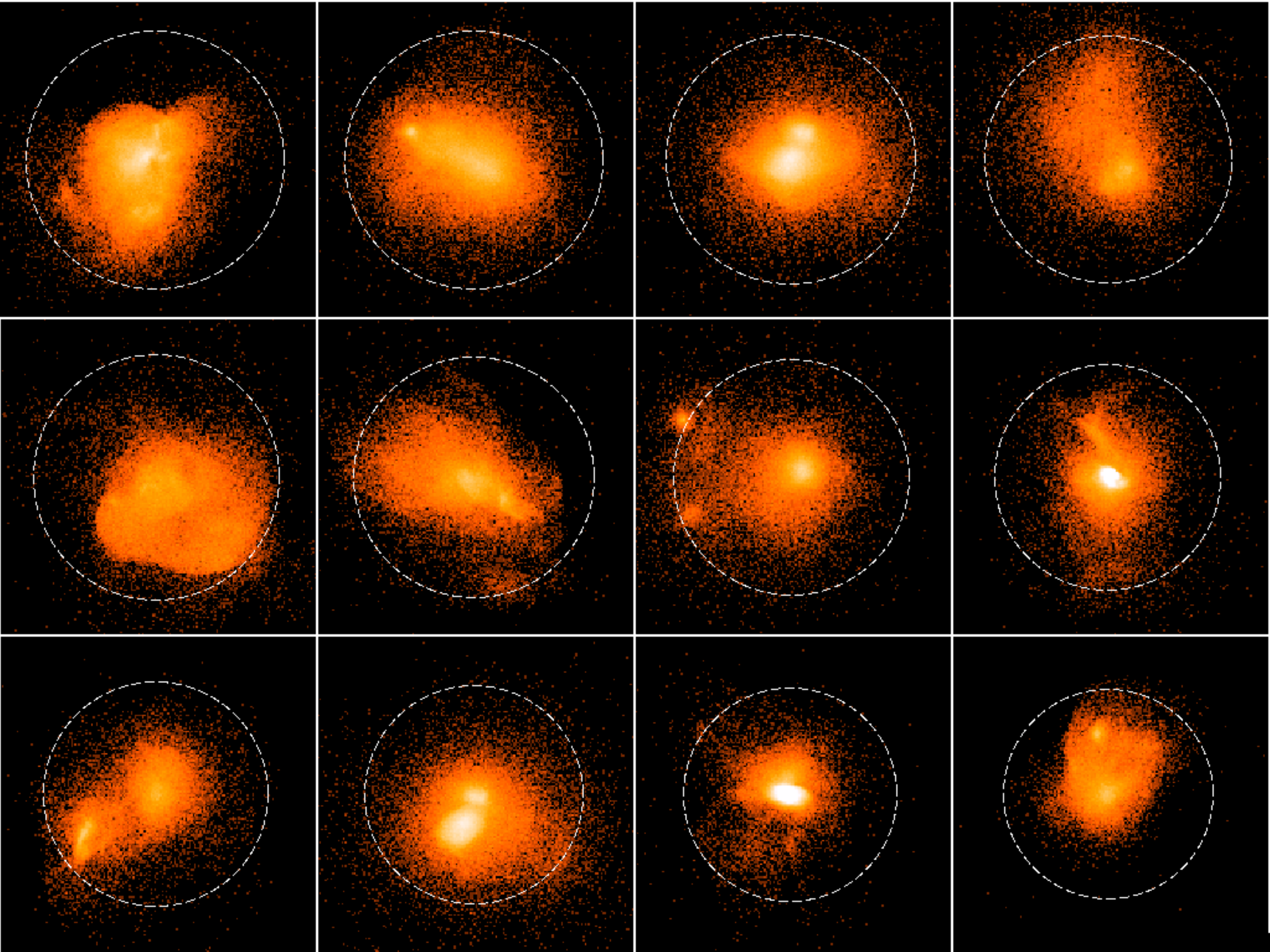}
\caption{ASTRO-H mock maps of the present-day clusters with $E_{\rm kin}/E_{\rm tot}<0.05$ (top panel) and $E_{\rm kin}/E_{\rm tot}>0.25$ (bottom panel). Each image is $2.8$~Mpc across. The white circles indicate $r_{500c}$ of each cluster. 
}
\label{fig2}
\end{center}
\end{figure}

\section{Results}
\label{sec:results}

\subsection{Predicting the Merger-induced Gas Flows in $\Lambda$CDM clusters}

We begin by considering the average gas velocity structure in the Bolshoi NR sample.
Fig.~\ref{fig1} shows the 1D mass-weighted gas velocity dispersion,
\begin{equation}
\sigma_{\rm mw} = \left( \frac{1}{3} \sum_{i=1}^3  \frac{ \sum_k m_k(v^i_k - v_{\rm pec}^i)^2 }{\sum_k m_k} \right)^{{1}/{2}},
\label{eq:sigma}
\end{equation}
where $m_k$ and $v^i_k$ are the gas mass and the $i$-th component of the gas velocity of the $k$-th cell, respectively, and $v_{\rm pec}^i$ is the $i$-th component of the cluster dark matter center-of-mass peculiar velocity within $r_{500c}$, as a function of $M_{500c}$. The best-fit power-law is shown by the middle solid line, with a slope fixed to the self-similar value of $1/3$:
\begin{equation}
\frac{\sigma_{\rm mw}(<r_{500c})}{(212\pm 3) \,{\rm km\,s^{-1}}} = \left( \frac{M_{500c}}{10^{14}h^{-1}M_{\odot}} \right )^{1/3}.
\label{eq1}
\end{equation}
The two surrounding solid lines indicate $\pm2\sigma$ around the best-fit line, showing considerable scatter in $\sigma_{\rm mw}$ at any given mass.  At $M_{500c}=10^{14}h^{-1}M_{\odot}$, the line-of-sight mass-weighted gas velocity dispersion ranges from $100-400$~km s$^{-1}$.  The red and blue points indicate clusters with the kinetic energy fraction of $E_{\rm kin}/E_{\rm tot}>0.25$ and $<0.05$, respectively, where the kinetic energy is defined as $E_{\rm kin}=\frac{3}{2}M_{\rm gas}\sigma^2_{\rm mw}(<r_{500c})$, and $E_{\rm tot}$ is the sum of gas kinetic and thermal energy. The dependence of $\sigma_{\rm mw}$ on the kinetic energy fraction suggests that the Doppler broadening of the Fe line detectable with ASTRO-H depends on the cluster dynamical state, and the magnitude of the broadening is expected to be larger for systems with larger kinetic energy fraction. Note that systems with $E_{\rm kin}/E_{\rm tot}>0.25$ are some of the most disturbed clusters in our sample, so large variations are expected for this population. For example, one red point that deviates significantly below the rest of the population is a merging cluster, in which $r_{500c}$ of the main cluster encompasses parts of the merging sub-cluster, but the gas velocity dispersion has not been affected significantly as the merging sub-cluster is still in its first passage. 

In Fig.~\ref{fig2}, we show ASTRO-H mock maps of the simulated clusters with $E_{\rm kin}/E_{\rm tot}<0.05$ and $E_{\rm kin}/E_{\rm tot}>0.25$.  These images show that systems with large kinetic energy fractions are morphologically disturbed, while those with small kinetic energy fractions appear relaxed with little substructure or asymmetries. This indicates that morphological properties of galaxy clusters, which can be measured using spatially-resolved \emph{Chandra} and \emph{XMM}-Newton X-ray observations, can serve as useful indicators for the level of internal gas motions in clusters. 

The best-fit slope of the $\sigma_{\rm mw}-M$ relation is $0.422\pm0.023$, steeper than the self-similar slope of $1/3$. This is driven in part by the presence of relaxed, low-mass clusters (indicated by blue points). These are groups of galaxies which formed at high-redshift and have not experienced significant mergers or mass accretion recently. In the hierarchical structure formation model, these early forming systems are common at low-masses, but are much rarer among massive clusters that are still forming today.  This produces an excess of low-mass groups with low gas velocity dispersion, and a number of them lie below the $-2\sigma$ line (indicated by the lower thin grey line). We note that this is analogous to the concentration-mass relation in the $\Lambda$CDM model, which also exhibits large scatter and dependence on formation time, which is responsible for the abundance of highly concentrated, early-forming low-mass groups \citep{wechsler_etal02,zhao_etal09}.

\begin{figure}[t]
\begin{center}
\epsscale{1.2}\plotone{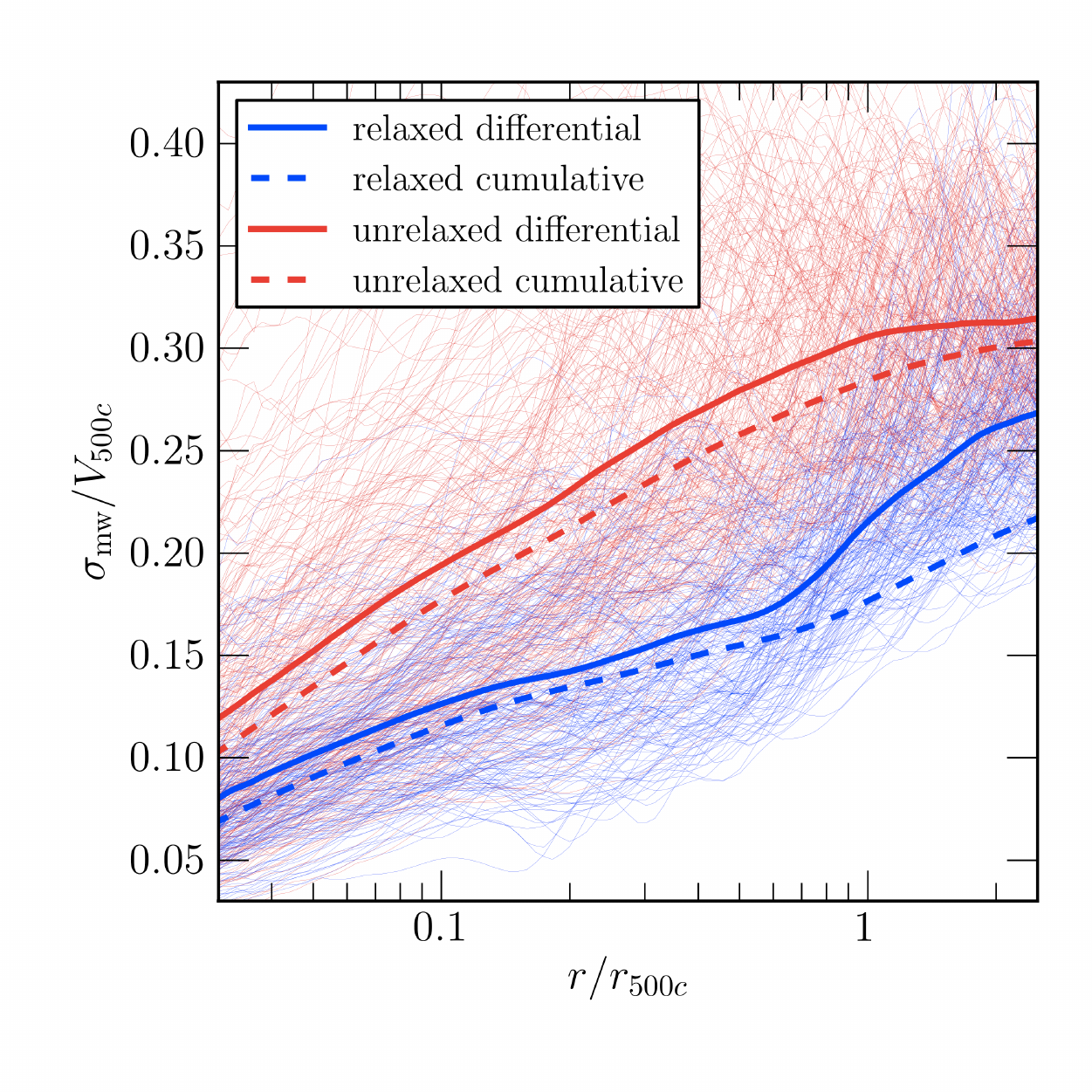}
\caption{One dimensional mass-weighted velocity dispersion of gas as a function of $r/r_{500c}$. The profile is normalized to $V_{500c}\equiv \sqrt{GM_{500c}/r_{500c}}$. Solid lines indicate the differential gas velocity dispersion $\sigma_{\rm mw}(r)$ in each annulus.  Dashed lines indicate the cumulate gas velocity dispersion $\sigma_{\rm mw}(<r)$ within a sphere with radius $r$.  Thick lines are median profiles for relaxed (blue with $E_{\rm kin}/E_{\rm tot}<0.1$) and unrelaxed clusters (red with $E_{\rm kin}/E_{\rm tot}>0.1$). Thin lines are individual cluster profiles.
}
\label{fig3}
\end{center}
\end{figure}

\subsection{Impact of Cluster Core Physics on Gas Velocity Structure}

Next, we investigate how the gas velocity dispersion changes with cluster-centric radius.  Fig.~\ref{fig3} shows the 1D mass-weighted velocity dispersion for the Bolshoi NR sample, normalized to the circular velocity $V_{500c}\equiv \sqrt{GM_{500c}/r_{500c}}$. We plot the 1D differential mass-weighted gas velocity dispersion $\sigma_{\rm mw}(r)$ in each radial shell $[r,r+dr]$ (solid lines) and the 1D cumulative mass-weighted gas velocity dispersion $\sigma_{\rm mw}(<r)$ within a sphere with radius $r$ (dashed lines), respectively.  We also compare the median profiles of relaxed (indicated by blue lines) and unrelaxed clusters (indicated by red lines).  The latter shows a large scatter because their morphology deviates significantly from spherical symmetry. Here we divide our sample along the median kinetic energy fraction of the sample, into relaxed and unrelaxed if their kinetic energy fraction within $r_{500c}$ is less than or greater than $0.1$, respectively. Our main finding is that the median gas velocity dispersion increases monotonically with radius.  It is therefore important to measure the gas velocity dispersion as a function of radius in order to measure the total kinetic energy of a galaxy cluster.

\begin{figure}[t]
\begin{center}
\epsscale{1.2}\plotone{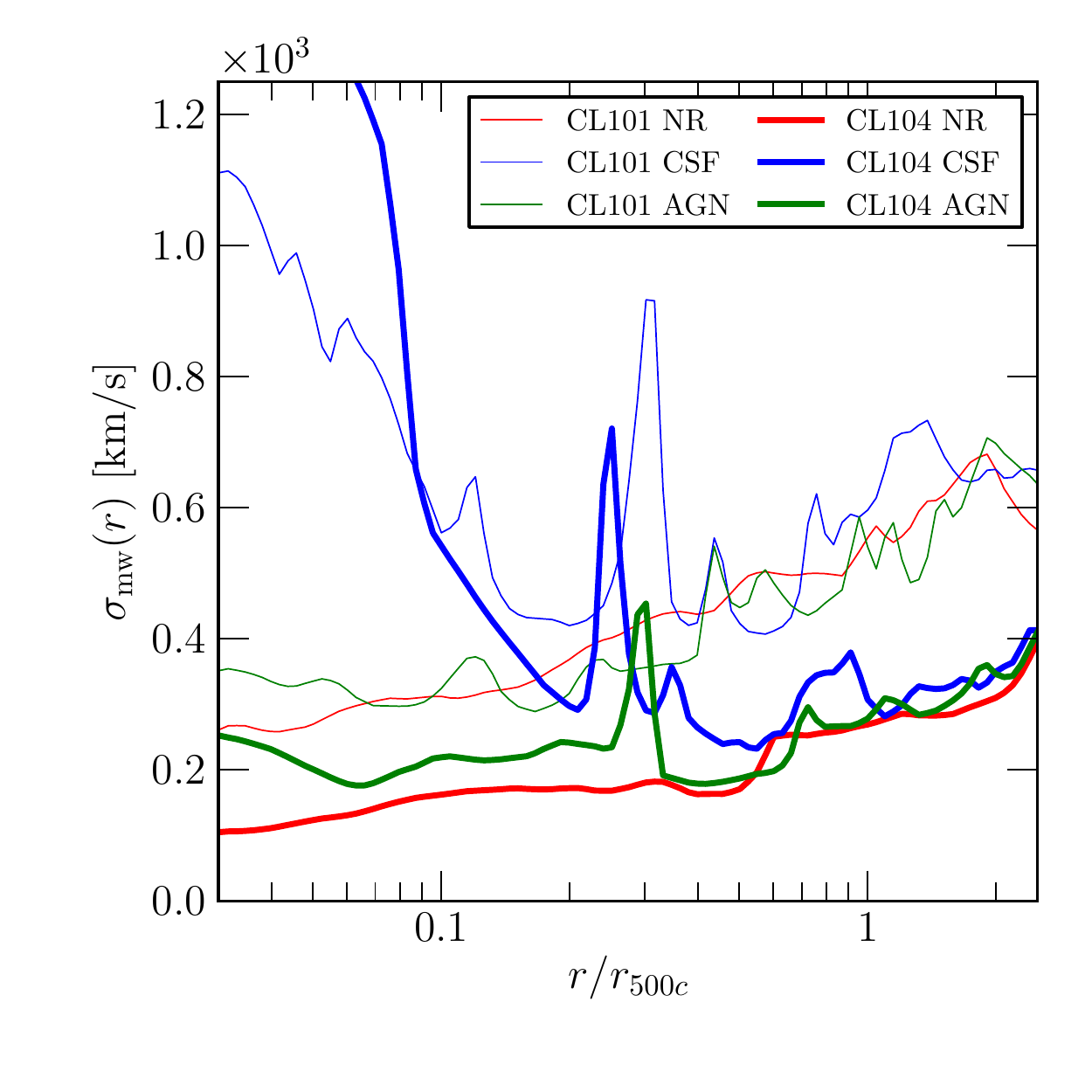}
\caption{One dimensional differential mass-weighted velocity dispersion of gas $\sigma_{\rm mw}(r)$ as a function of $r/r_{500c}$ for simulations with varying input physics.  Red, blue, and green lines indicate runs with non-radiative (NR), cooling and star formation (CSF), and feedback from active galactic nuclei (CSF+AGN). Thin and thick lines indicate CL101 (unrelaxed) and CL104 (relaxed), respectively. 
}
\label{fig4}
\end{center}
\end{figure}

\begin{figure}[htbp]
\begin{center}
\epsscale{0.92}\plotone{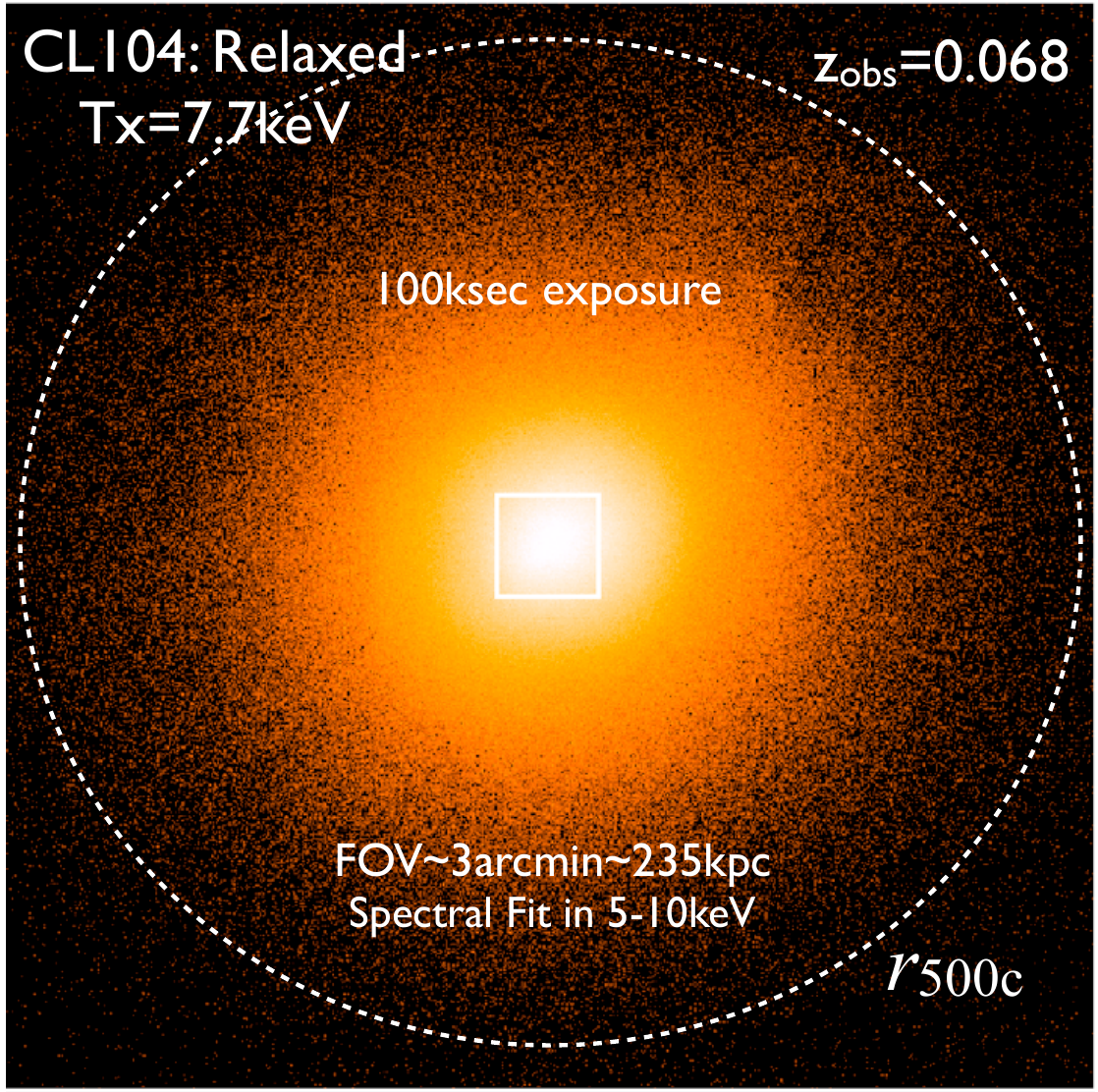}
\epsscale{1.2}\plotone{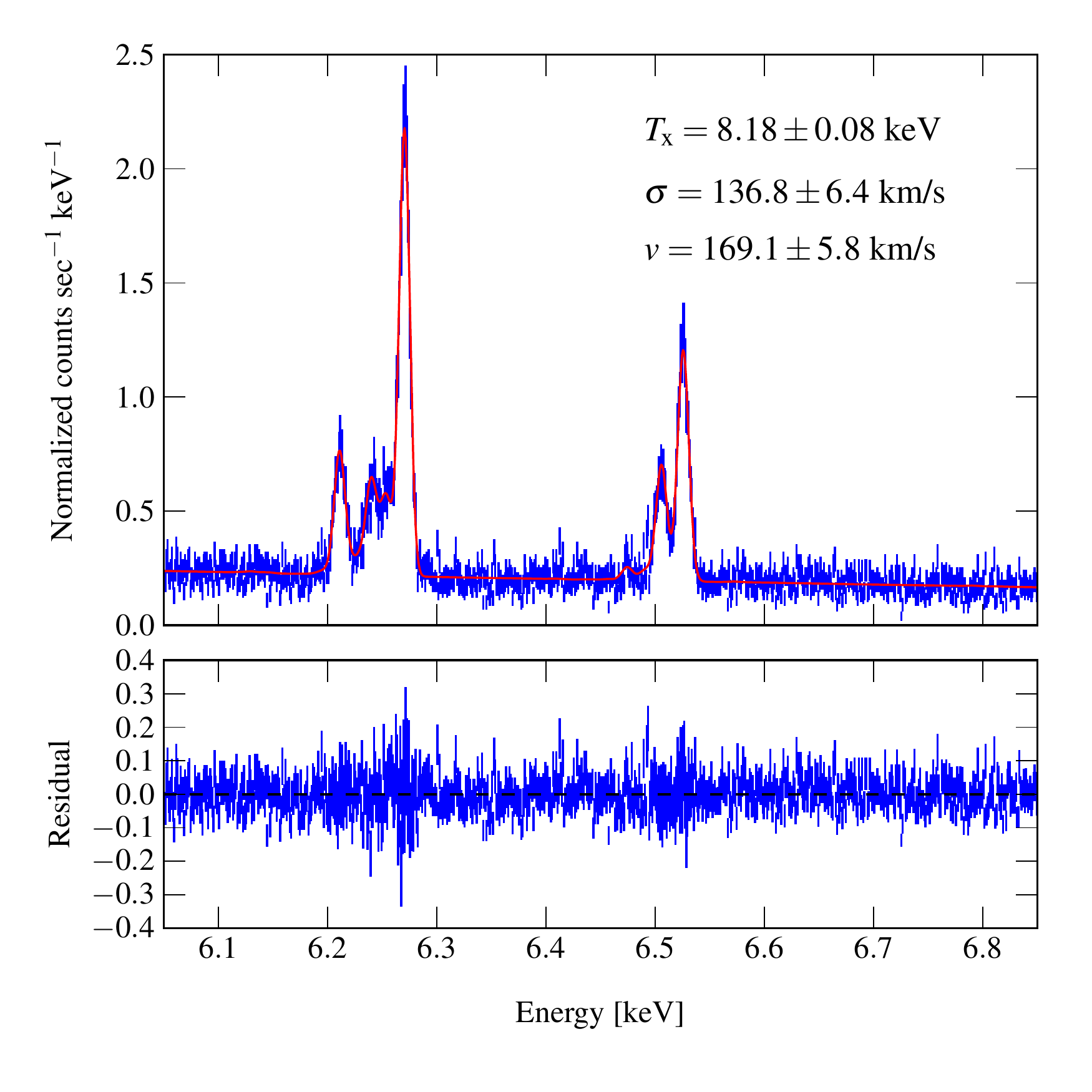}
\caption{Mock ASTRO-H analysis of a relaxed cluster CL104 for the NR run. {\it Top panel:} Mock ASTRO-H image in $5-10$~keV band. The region shown is about 2.6~Mpc across, and the dotted circle indicates $r_{500c}$. {\it Bottom panel:} Mock ASTRO-H spectra of the central region with a $100$~ksec exposure. The red line shows the best-fit spectrum. }
\label{fig5}
\end{center}
\end{figure}

However, the cluster gas velocity structure is expected to depend sensitively on cluster physics, especially in the central regions where gas cooling and energy injection from supernova and AGN can significantly influence both the thermodynamic and velocity structure of the ICM.  To assess these effects, in Fig.~\ref{fig4} we show the 1D differential mass-weighted velocity dispersion of gas $\sigma_{\rm mw}(r/r_{500c})$ for two simulated clusters (CL104 and CL101), each performed with varying input gas physics: non-radiative (NR); cooling and star formation (CSF); and cooling, star formation, plus thermal AGN feedback (CSF+AGN).  Gas cooling and star formation have a significant effect on the velocity structure of the ICM, especially in the central regions ($r\lesssim 0.3\,r_{500c}$). Note, however, that the CSF runs suffer from the well-known ``overcooling" problem, which leads to an over-production of stars as well as a high-level of star formation at late times, making the simulated central galaxy considerably bluer than those observed.  The gas velocities near the cluster center in the CSF runs are dominated by strong coherent rotation as the gas ``overcools" to form a rotationally supported disk, making the gas velocity dispersion significantly larger in the cluster core \citep{lau_etal11}.\footnote{Note that our definition of $\sigma_{\rm mw}$ (Eq.~\ref{eq:sigma}) includes both mean {\em and} random velocity components.} This effect is more pronounced in the relaxed cluster (CL104) in which the gas rotation has not been disrupted by mergers.  Such strong rotation in the CSF run induces a significant increase in the ellipticity of the central X-ray surface brightness, which has been ruled out by \emph{Chandra} observations \citep{fang_etal09, lau_etal12}. Our CSF+AGN runs are more realistic; for example, the inclusion of AGN helps regulate gas cooling and better reproduces the mass and color of the bright central galaxy as well as the observed stellar fraction (Avestruz et al., in prep).  As shown in Fig.~\ref{fig5}, the gas velocity dispersion profiles in the CSF+AGN runs are similar to those of the non-radiative runs at all radii.  In the relaxed cluster (CL104), we find enhancement in gas velocity at the level of $30-50$~km~s$^{-1}$ within $r<0.3\,r_{500c}$, while the difference is not discernible in the unrelaxed system (CL101) in which the presence of significant (of order $300$~km~s$^{-1}$) merger-induced gas flows dominate over the gas flows generated by the cluster core physics. Note that mergers could drive gas feeding into the supermassive black hole in the central cluster galaxy, but the kinetic energy injected by AGN should be a few orders of magnitude smaller than that induced by mergers.

As shown in Fig.~\ref{fig4}, the differential gas velocity dispersion at $r_{500c}$ is insensitive to the included cluster physics. This means that ASTRO-H measurements of gas motions in the outskirts of galaxy clusters could be used to estimate the non-thermal pressure support and may even provide correction factors for estimates of the mass assuming hydrostatic equilibrium \citep[][Lau et al. in press.]{nelson_etal12}. The cumulative gas velocity dispersion $\sigma_{\rm mw}(<r_{500c})$, on the other hand, is more susceptible to baryonic physics in the cluster core. In the CSF+AGN simulation, $\sigma_{\rm mw}$ is modified by -5\% for CL101 and +16\% for CL104, compared to the NR simulation. As discussed above, the effect of cluster core physics is likely significantly overestimated, and the cumulative gas velocity dispersion is larger by +190\% in CL104 and +16\% in CL101 with respect to the NR run, indicating that the ASTRO-H measurements might shed light on the impact of baryon physics on the gas velocity structures. 

\begin{figure}[htbp]
\begin{center}
\epsscale{0.92}\plotone{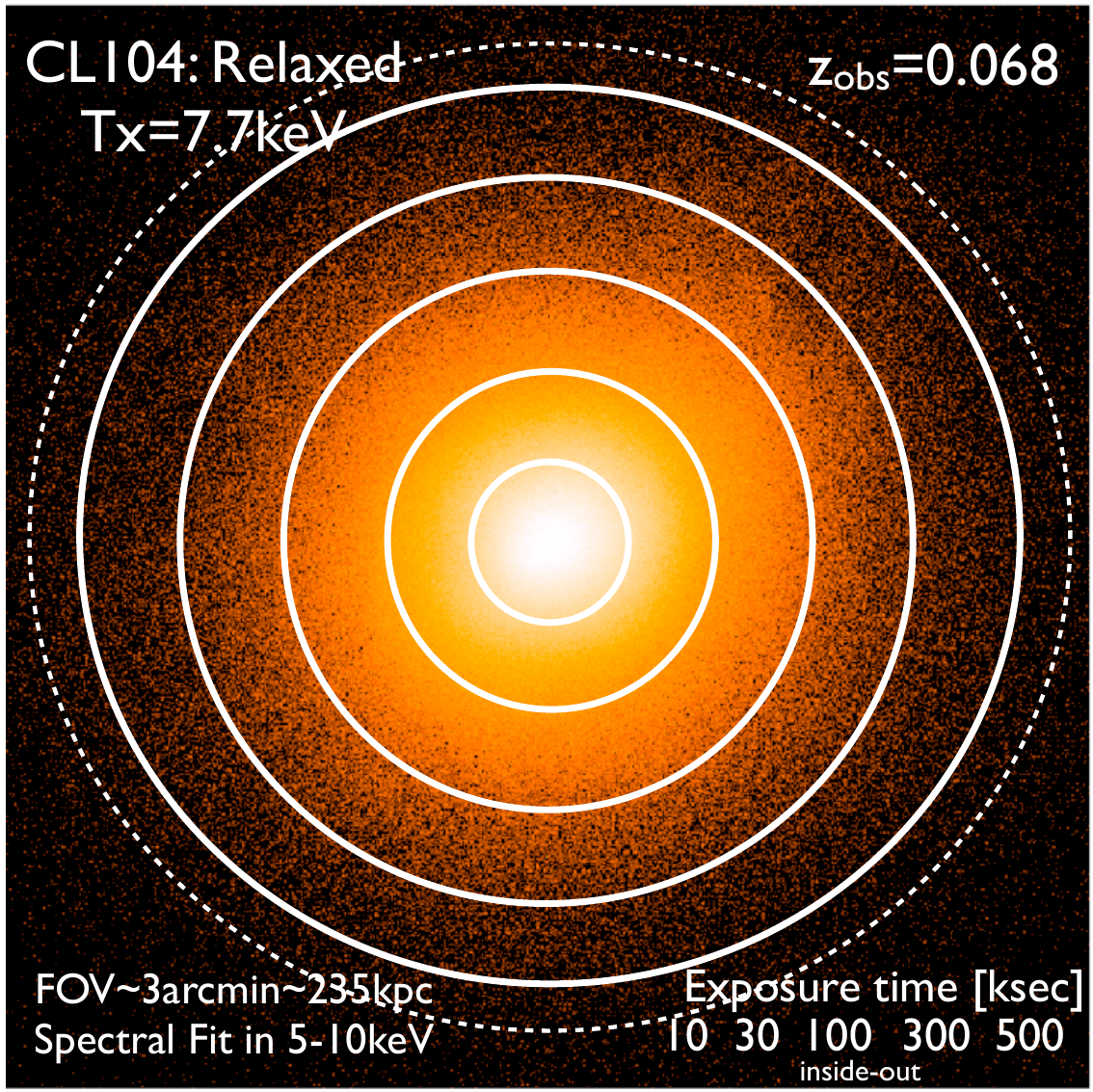}
\epsscale{1.2}\plotone{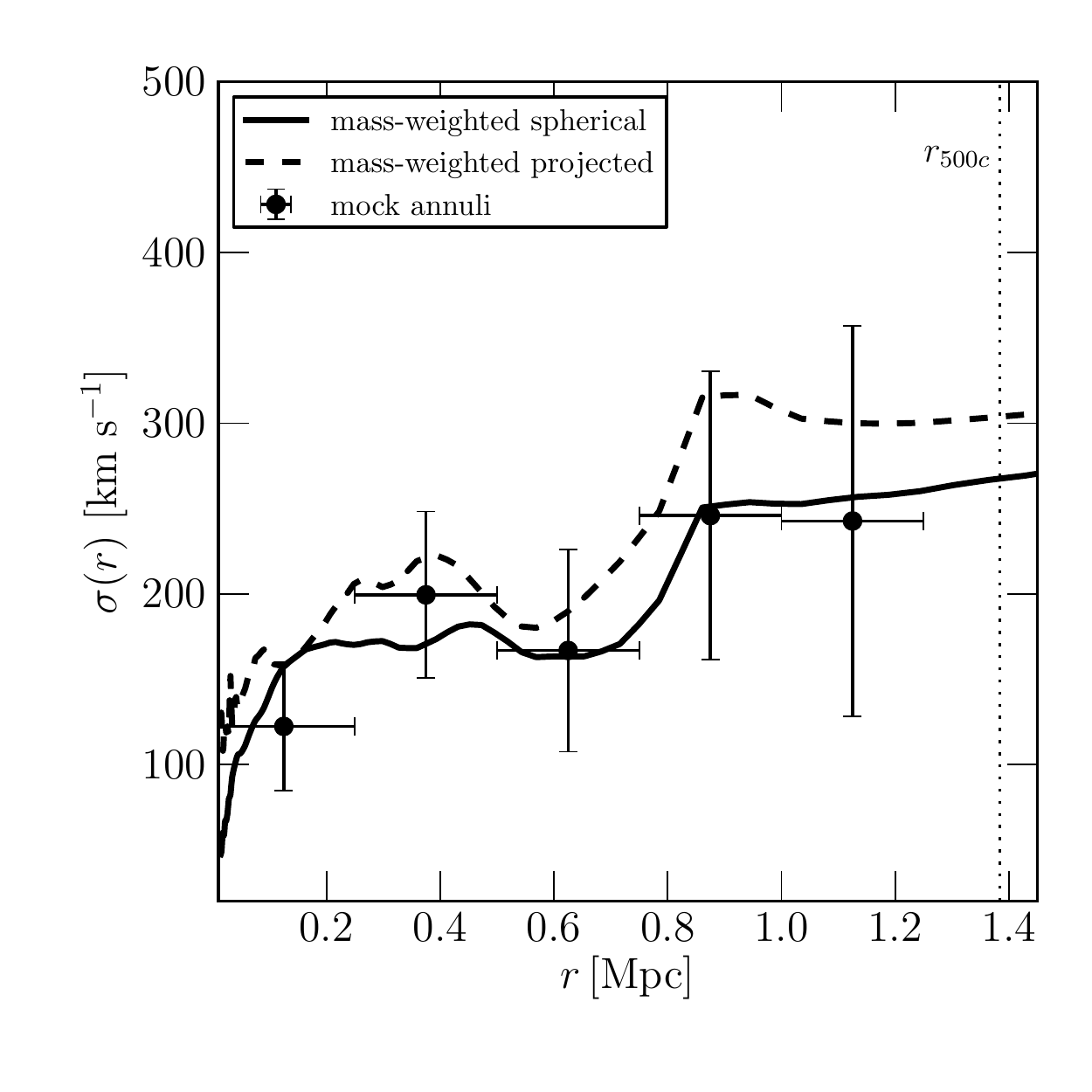}
\caption{Mock ASTRO-H analysis of a relaxed cluster CL104 for the NR run. {\it Top panel:} Mock ASTRO-H image in a $5-10$~keV band. The region shown is about $2.6$~Mpc across, and the dotted circle indicates $r_{500c}$. {\it Bottom panel:} ASTRO-H measurements of the gas velocity dispersion as a function of radius with $10$, $30$, $100$, $300$, $500$~ksec exposures. The solid black line is the differential mass-weighted profile in a spherical shell. The dashed black line is the projected differential mass-weighted gas velocity dispersion profile. 
}
\label{fig6}
\end{center}
\end{figure}

\subsection{Measuring Gas Motions in the Outskirts of Relaxed Galaxy Clusters with ASTRO-H} 

In this section, we explore the detectability of the velocity structure in the ICM with the upcoming ASTRO-H mission.  Fig.~\ref{fig5} shows the mock ASTRO-H surface brightness map and spectrum of the relaxed cluster CL104 in the NR run at the fiducial redshift $z = 0.068$. The top panel shows the mock ASTRO-H surface brightness map in a $5-10$ keV band, showing a region of $2.6$~Mpc encompassing $r_{500c}$ of this cluster. The bottom panel shows the mock ASTRO-H spectral analysis of data extracted from a 100~ksec exposure of the central region, indicated by the white box in the top panel. A single component model spectrum (shown in red) provides a good fit to the spectrum extracted from this region (shown in blue). The best-fit velocity dispersion is $137\pm 6$~km~s$^{-1}$, and the best-fit peculiar velocity is $169\pm 6$~km~s$^{-1}$, demonstrating the power of ASTRO-H for studying the gas velocity structure in the central regions where we can expect to obtain large X-ray photon counts. 

Extending this analysis to large radii would be challenging for ASTRO-H, as the surface brightness declines rapidly toward the cluster outskirts, but it is still possible with very deep ASTRO-H observations of nearby relaxed clusters as we demonstrate here.  Figure~\ref{fig6} shows an example deep ASTRO-H observational program to measure the gas velocity dispersion as a function of radius out to $r\sim r_{500c}$.  From inside out, we allocated $10$, $30$, $100$, $300$, and $500$~ksec exposures, such that the measurement errors are roughly comparable in size, shown in the top panel.  The total integration time is $940$~ksec.  A detailed characterization of the gas velocity profile out to $r\approx r_{500c}$ will require of order 1~Msec of ASTRO-H time, with a significant time spent on the outermost radial bins.  The data points in the bottom panel show the gas velocity dispersion extracted from the spectral fitting of the ASTRO-H spectra extracted from the radial shell, indicated by the horizontal errorbar. The measured gas velocity is in good agreement with the 3D cumulative mass-weighted velocity dispersion profile, and it is slightly ($\sim 30-50$~~km~s$^{-1}$) smaller than the differential mass-weighted velocity dispersion within the projected radial bin. The difference is partly due to the fact that the measured velocity is spectral-weighted.  This leads to an underestimate of the gas velocity dispersion relative to the mass-weighted ones, because the spectral weighting gives more weight to the inner regions where the gas density is higher but the gas velocity is smaller. 

\begin{figure}[htbp]
\begin{center}
\epsscale{0.92}\plotone{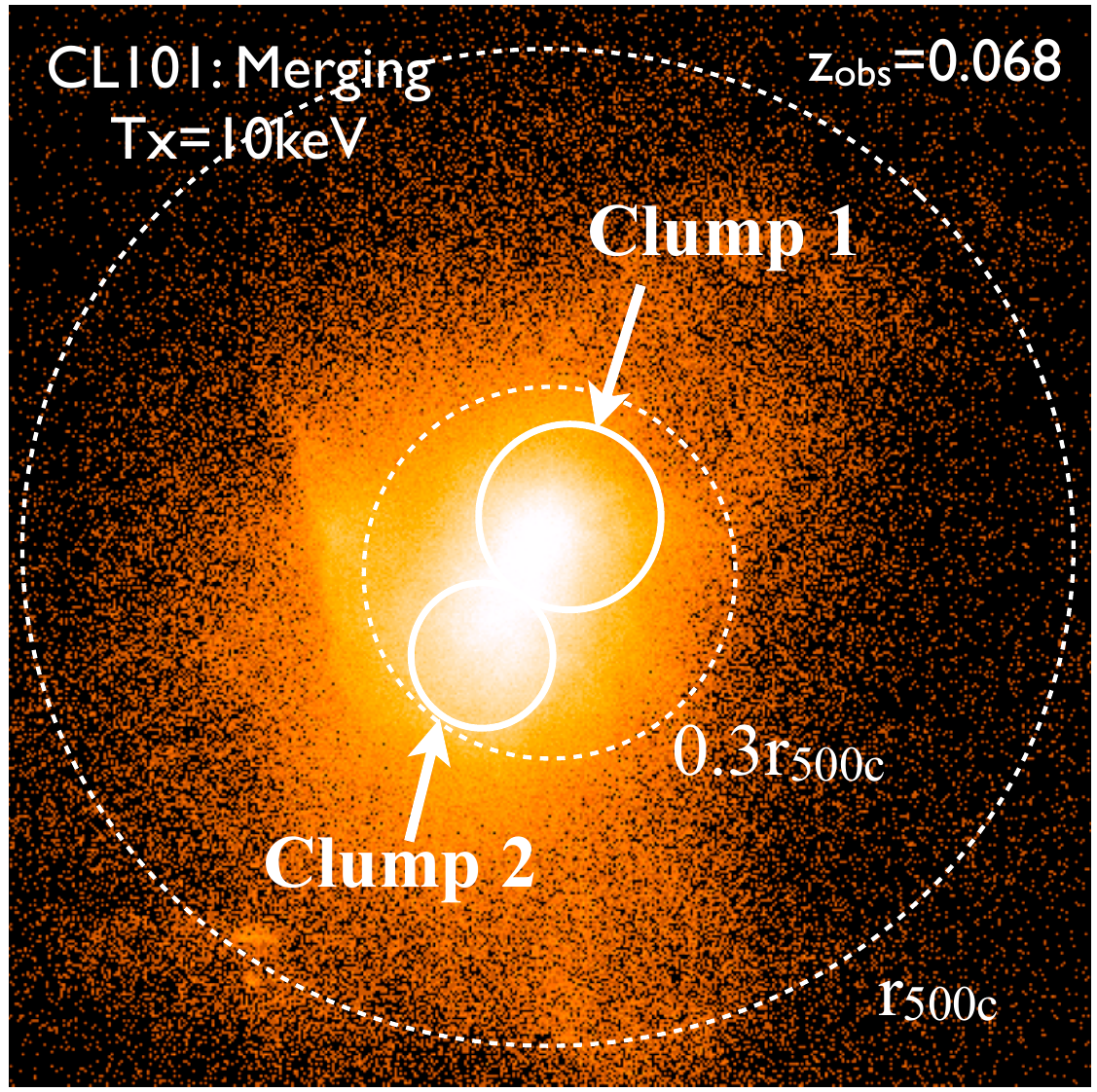}
\epsscale{1.2}\plotone{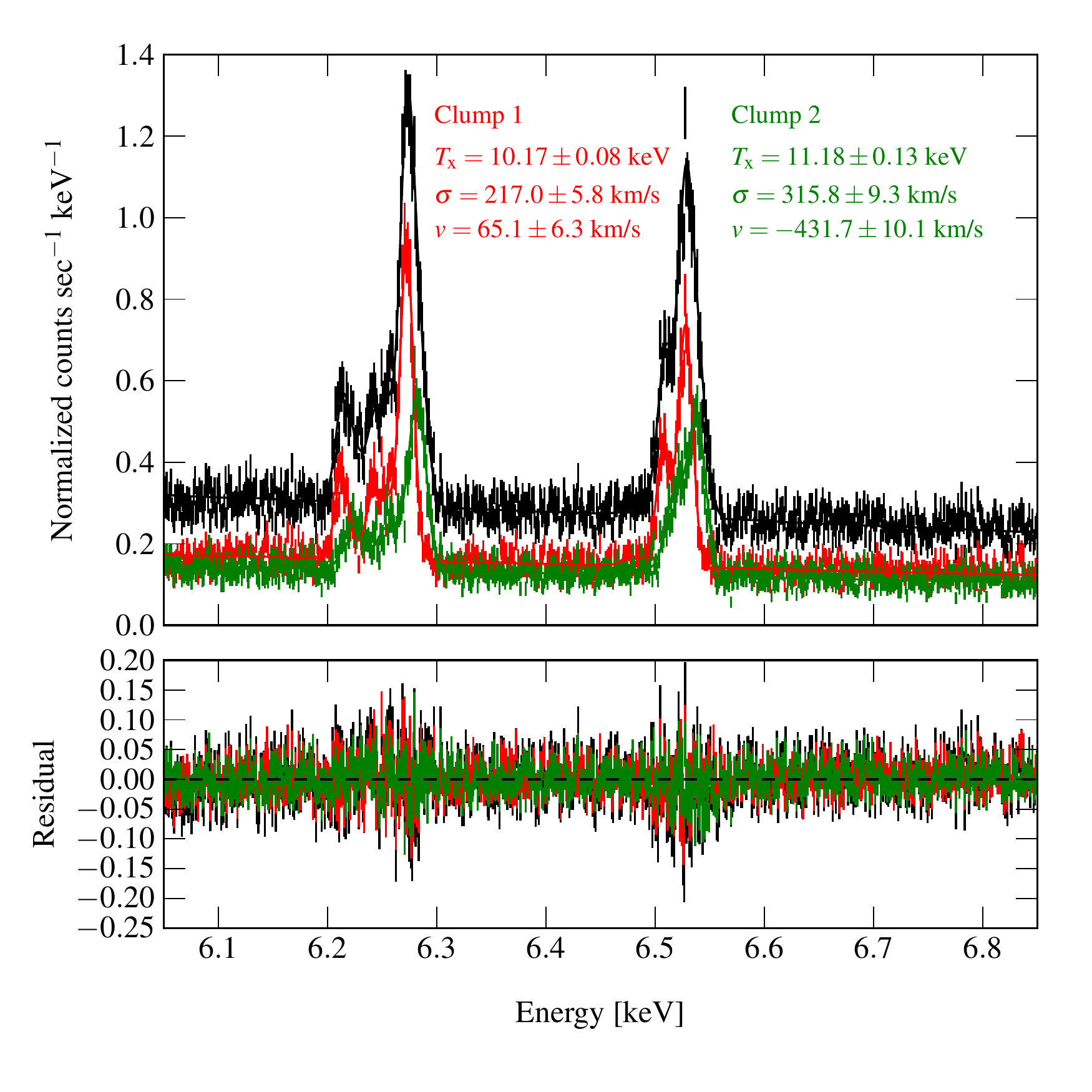}
\caption{Mock ASTRO-H analysis of a merging cluster CL101 for the NR run. {\it Top panel:} Mock ASTRO-H image in a $5-10$~keV band with a $300$~ksec exposure. The region shown is 2.6~Mpc across, and the inner and outer dotted circles indicate $0.3\times r_{500c}$ and $r_{500c}$, respectively. We extract spectra from regions labeled clump 1 and clump 2. {\it Bottom panel:} Individual mock ASTRO-H spectra and the best-fit for clump 1 (red) and 2 (green). The fit of the combined spectrum of clumps 1 and 2 (black line) is the sum of the individual fits. }
\label{fig7}
\end{center}
\end{figure}

\subsection{Studying Substructures in Merging Galaxy Clusters with ASTRO-H}

ASTRO-H analysis of merging galaxy clusters is more involved, requiring multiple spectral components. Figure~\ref{fig7} shows one such example: the mock ASTRO-H analysis of cluster CL101 in the NR run, which consists of two merging sub-clusters 1 and 2.  A single component spectral model underestimates the cluster peculiar velocity (due to the cancelation of peculiar velocities associated with two merging components with opposite signs) and overestimate of the gas velocity dispersion.  The magnitude of these biases depends on the orientation of the merger axis and the line-of-sight, and the biases are the largest when the two merging sub-clusters are viewed along the line-of-sight.  A two component fit provides a much better description of the data, yielding accurate measurements of the peculiar velocity and gas velocity dispersion for each component.  Spectral analysis based on a Bayesian Gaussian Mixture model may also help disentangle multiple velocity components robustly \citep{shang_oh12}. This example illustrates that the ASTRO-H spectrum contains rich information about substructures in the ICM, enabling us to probe structures along the line of sight. The ASTRO-H spectrum will be highly complementary to studies of gas clumps in spatially resolved 2D X-ray imaging with \emph{Chandra} and \emph{XMM}-Newton observations. 

\section{Summary \& Discussion}
\label{sec:summary}

In this work, we investigate internal gas motions within the virialized regions of galaxy clusters and their detectability with the upcoming ASTRO-H mission. Using a large, volume-limited sample of simulated galaxy clusters and groups in the $\Lambda$CDM model, we characterize the range of internal gas motions as a function of mass and dynamical state. We find that the gas velocity dispersion is larger in more massive clusters, but exhibits large scatter due to the diversity in their dynamical states.  We show that systems with large gas motions are generally morphological disturbed, while early forming, relaxed groups show a considerably smaller level of merger-induced gas motions. This suggests that morphological properties of galaxy clusters that can be measured using spatially-resolved \emph{Chandra} and \emph{XMM}-Newton X-ray observations should be good indicators for the level of internal gas motions in clusters. 

We show that the gas velocity structure changes with cluster-centric radius.  We find that the gas velocity dispersion on average increases monotonically with radius in clusters, because cluster outskirts are the regions where materials are actively accreting and the relaxation time becomes progressively longer.  This means that it is important to measure the gas velocity dispersion as a function of radius in order to constrain the kinetic energy and hence the total energy budget of galaxy clusters as a whole.  Baryonic physics, such as gas cooling and energy injection from supernovae and AGN, affect the velocity structure of the gas in inner regions \citep[see also][]{biffi_etal11,dubois_etal11,vazza_etal13}. Our simulations suggest that the effects of cluster core physics is confined within $r\lesssim 0.3\,r_{500c}$ for massive clusters. We also show that simulations without gas cooling and star formation (without AGN) significantly overestimate the gas velocity in the central regions of clusters.  The simulations with with AGN, on the other hand, are more similar to those in the non-radiative run, because AGN provide additional energy necessary to regulate gas flow induced by gas cooling and star formation. We find that the inclusion of AGN increases gas velocity dispersion on the order of $30-50$~km~s$^{-1}$ in the central regions. This effect should be discernible in relaxed clusters, where the merger induced gas motions are smaller, but it is more difficult to see in dynamically active clusters in which the merger-induced gas motions are on the order of $300-500$~km~s$^{-1}$. 

Finally, we show that deep ASTRO-H observations can accurately measure the gas velocity dispersion to $r\approx r_{500c}$ in nearby relaxed clusters.  Although the Doppler broadening should be larger for merging clusters, ASTRO-H spectral analysis would be more involved for such objects, requiring multiple spectral components.  However, we also point out that the multi-component spectral fitting is a powerful approach for studying substructures in velocity space, which would be highly complementary to studies of gas clumps with spatially resolved \emph{Chandra} and \emph{XMM}-Newton observations.  Our analyses indicate that ASTRO-H should be capable of constraining the missing energy associated with internal gas motions in clusters and enable unique studies of their substructures. 

There are several caveats that must be kept in mind when interpreting our results.  First, plasma instabilities, such as magneto-thermal instability (MTI), may generate additional turbulent motions and contribute up to $\sim 40\%$ of extra non-thermal pressure and $\sim 20\%$ increase in the one dimensional differential gas velocity dispersion at $r=r_{500c}$ (\citealt{parrish_etal12}, but see also \citealt{ruszkowski_etal11}). To our knowledge, this is the largest systematic uncertainty in our predicted gas velocity dispersion, as our simulations do not include magnetic field and related plasma effects.  With deep ASTRO-H cluster observations, the accuracy of velocity measurements may reach $20$~km~s$^{-1}$, which is comparable to the increase in velocity dispersion due to MTI for a massive relaxed cluster.  Therefore, ASTRO-H might be able to distinguish models with and without MTI.  Furthermore, there are additional complications in real ASTRO-H spectral analysis that might introduce additional observational uncertainties. For example, the spectral analysis of cluster outskirts may be complicated by the difficulty in disentangling the low surface brightness signal associated cluster emission from large background and foreground emissions. Our mock ASTRO-H analysis also ignored the effect of resonant scattering, which might introduce additional uncertainties in the gas velocity measurements. 

Over the next several years, the velocity structure of the ICM will become an important and active area of research. Observationally, ASTRO-H will make a first direct detection of the internal gas motions in galaxy clusters. However, these measurements are likely limited to a handful of nearby massive galaxy clusters. Next generation X-ray missions, such as \emph{Athena}+\footnote{\url{http://www.the-athena-x-ray-observatory.eu}} and SMART-X\footnote{\url{http://hea-www.cfa.harvard.edu/SMARTX}}, are necessary to extend such study for a cosmologically representative sample as well as higher-redshift and lower mass groups and to provide fuller insights into the missing energy problem in galaxy clusters. Theoretically, there are still a number of important issues that must be addressed before we could interpret ASTRO-H data in more detail.  For example, the internal gas motions consist of the combination of bulk and turbulent gas flows, but their relative importance still remains unclear and must be understood. Aforementioned plasma instabilities may also give rise to additional turbulent gas motions, but no work to date has characterized the magnitude of this effect using cosmological simulations. Future work should thus focus on characterizing bulk and turbulent gas motions and their relative importance, understanding their physical origin, and assessing the importance of plasma effects. 

\acknowledgments 
We thank Tetsu Kitayama, Maxim Markevitch, Naomi Ota, Cien Shang, Andrew Szymkowiak and Irina Zhuravleva for useful discussions and/or comments on the manuscript, and Anatoly Klypin for providing initial conditions for the Bolshoi simulation. We acknowledge support from NSF grants AST-1009811 and OCI-0904484, NASA ATP grant NNX11AE07G, NASA Chandra Theory grant GO213004B, Research Corporation, and by Yale University. CA acknowledges support from NSF Graduate Student Research Fellowship and Alan D. Bromley Fellowship from the Yale University. This work was also supported in part by the facilities and staff of the Yale University Faculty of Arts and Sciences High Performance Computing Center. 

\pdfbookmark{REFERENCES}{references}
\bibliographystyle{apj_ads}\bibliography{ms}

\end{document}